# A Multiphilic Descriptor for Chemical Reactivity and Selectivity

**J. Padmanabhan[1,2], R. Parthasarathi[2], M. Elango[2], V. Subramanian[2,*], B. S. Krishnamoorthy[1,3], S. Gutierrez-Oliva[4], A. Toro-Labbé[4,*], D. R. Roy[1] and P. K. Chattaraj[1,*]**

[1]*Department of Chemistry, Indian Institute of Technology, Kharagpur 721302, India.*
[2]*Chemical Laboratory, Central Leather Research Institute, Adyar, Chennai 600 020, India.*
[3]*School of Chemistry, Bharathidasan University, Tiruchirappalli-620 024, India.*
[4]*Laboratorio de Química Teórica Computacional (QTC), Facultad de Química, Pontificia Universidad Católica de Chile, Casilla 306, Correo 22, Santiago, Chile.*

**Abstract**

In line with the local philicity concept proposed by Chattaraj et al. (Chattaraj, P. K.; Maiti, B.; Sarkar, U. *J. Phys. Chem. A.* **2003**, *107*, 4973) and a dual descriptor derived by Toro-Labbé and coworkers (Morell, C.; Grand, A.; Toro-Labbé, A. *J. Phys. Chem. A.* **2005**, *109,* 205), we propose a multiphilic descriptor. It is defined as the difference between nucleophilic ($\omega_k^+$) and electrophilic ($\omega_k^-$) condensed philicity functions. This descriptor is capable of simultaneously explaining the nucleophilicity and electrophilicity of the given atomic sites in the molecule. Variation of these quantities along the path of a soft reaction is also analyzed. Predictive ability of this descriptor has been successfully tested on the selected systems and reactions. Corresponding force profiles are also analyzed in some representative cases. Also, to study the intra- and intermolecular reactivities another related descriptor namely, the nucleophilicity excess ($\Delta\omega_g^{\mp}$) for a nucleophile, over the electrophilicity in it has been defined and tested on all-metal aromatic compounds.

*Authors for correspondence:
E-mail: subuchem@hotmail.com, atola@puc.cl, pkc@chem.iitkgp.ernet.in,

## 1. Introduction

The understanding of chemical reactivity and site selectivity of the molecular systems has been effectively handled by the conceptual density functional theory (DFT).[1] Chemical potential, global hardness, global softness, electronegativity and electrophilicity are global reactivity descriptors, highly successful in predicting global chemical reactivity trends. Fukui function (FF) and local softness are extensively applied to probe the local reactivity and site selectivity. The formal definitions of all these descriptors and working equations for their computation have been described. [1-4] Various applications of both global and local reactivity descriptors in the context of chemical reactivity and site selectivity have been reviewed in detail.[3]

Parr et al. introduced the concept of Electrophilicity (ω) as a global reactivity index similar to the chemical hardness and chemical potential. [5] This new reactivity index measures the stabilization in energy when the system acquires an additional electronic charge ΔN from the environment. The electrophilicity is defined as

$$\omega = \mu^2 / 2\eta \tag{1}$$

In Eq. (1), $\mu \approx -(I+A)/2$ and $\eta \approx (I-A)/2$ are the electronic chemical potential and the chemical hardness of the ground state of atoms and molecules, respectively, approximated in terms of the vertical ionization potential (I) and electron affinity (A). The electrophilicity is a descriptor of reactivity that allows a quantitative classification of the global electrophilic nature of a molecule within a relative scale. [5]

Fukui Function (FF) [6] is one of the widely used local density functional descriptors to model chemical reactivity and site selectivity and is defined as the derivative of the electron density $\rho(\vec{r})$ with respect to the total number of electrons $N$ in the system, at constant external potential $v(\vec{r})$ acting on an electron due to all the nuclei in the system

$$f(\vec{r}) = \left[\delta\mu/\delta v(\vec{r})\right]_N = \left[\partial\rho(\vec{r})/\partial N\right]_{v(\vec{r})}. \quad (2)$$

The condensed FF are calculated using the procedure proposed by Yang and Mortier,[7] based on a finite difference method

$$f_k^+ = q_k(N+1) - q_k(N) \quad \text{for nucleophilic attack} \quad (3a)$$

$$f_k^- = q_k(N) - q_k(N-1) \quad \text{for electrophilic attack} \quad (3b)$$

$$f_k^o = \left[q_k(N+1) - q_k(N-1)\right]/2 \quad \text{for radical attack} \quad (3c)$$

where $q_k$ is the electronic population of atom k in a molecule.

Chattaraj et al.[8] have introduced the concept of generalized philicity. It contains almost all information about hitherto known different global and local reactivity and selectivity descriptors, in addition to the information regarding electrophilic/nucleophilic power of a given atomic site in a molecule. It is possible to define a local quantity called philicity associated with a site k in a molecule with the help of the corresponding condensed- to- atom variants of FF, $f_k^\alpha$ as

$$\omega_k^\alpha = \omega f_k^\alpha \quad (4)$$

where ($\alpha$= +, - and 0) represents local philic quantities describing nucleophilic, electrophilic and radical attacks respectively. Eq. (4) predicts that the most electrophilic site in a molecule is the one providing the maximum value of $\omega_k^+$. When two molecules react, which one will act as an electrophile (nucleophile) will depend on, which has a higher (lower) electrophilicity index. This global trend originates from the local behavior of the molecules or precisely at the atomic site(s) that is(are) prone to electrophilic (nucleophilic) attack. Recently the usefulness of electrophilicity index in elucidating the toxicity of polychlorinated biphenyls, benzidine and chlorophenol has been assessed in detail. [9-11]

In addition to the knowledge of global softness (S), which is the inverse of hardness, [12] different local softnesses [13] used to describe the reactivity of atoms in molecule, can be defined as

$$s_k^{\alpha} = Sf_k^{\alpha} \qquad (5)$$

where ($\alpha$= +, - and 0) represents local softness quantities describing nucleophilic, electrophilic and radical attacks respectively. Based on local softness, relative nucleophilicity ($s_k^- / s_k^+$) and relative electrophilicity ($s_k^+ / s_k^-$) indices have also been defined and their usefulness to predict reactive sites also been addressed to.[14] It has been established that the quantum chemical model selected to derive wave function; population scheme used to obtain the partial charges and basis set employed in the molecular orbital calculations are important parameters, which significantly influence the FF values. [15-18]

The condensed philicity summed over a group of relevant atoms is defined as the "group philicity". It can be expressed as[19]

$$\omega_g^{\alpha} = \sum_{k=1}^{n} \omega_k^{\alpha} \qquad (6)$$

where n is the number of atoms coordinated to the reactive atom, $\omega_k^{\alpha}$ is the local electrophilicity of the atom k, and $\omega_g^{\alpha}$ is the group philicity obtained by adding the local philicity of the nearby bonded atoms. In this study[19] the group nucleophilicity index ($\omega_g^+$) of the selected systems is used to compare the chemical reactivity trends.

Toro-Labbé et al[20] have recently proposed a dual descriptor ($\Delta f(r)$), which is defined as the difference between the nucleophilic and electrophilic Fukui functions and is given by,

$$\Delta f(r) = [\, (f^{+}(r) - (f^{-}(r) \,] \quad (7)$$

If $\Delta f(r) > 0$, then the site is favored for a nucleophilic attack, whereas if $\Delta f(r) < 0$, then the site may be favored for an electrophilic attack. The associated dual local softness have also been defined as,[19]

$$\Delta s_k = S\,(f_k^{+} - f_k^{-}) = (s_k^{+} - s_k^{-}) \quad (8)$$

It is defined as the condensed version of $\Delta f(r)$ multiplied by the molecular softness S.

**2. Multiphilic Descriptor**

In the light of the local philicity concept proposed by Chattaraj et al.[8] and the dual descriptor derived by Toro-Labbé and coworkers,[20] we propose a multiphilic descriptor using the unified philicity concept, which can concurrently characterize both nucleophilic and electrophilic nature of a chemical species. It is defined as the difference between the nucleophilic and electrophilic condensed philicity functions. It is an index of selectivity towards nucleophilic attack, which can as well characterize an electrophilic attack and is given by,[21]

$$\Delta \omega_k = [\omega_k^{+} - \omega_k^{-}\,] = \omega\,[\Delta f_k] \quad (9)$$

where $\Delta f_k$ is the condensed-to-atom variant-$k$ of $\Delta f(r)$ (eq 7). If $\Delta\omega_k > 0$, then the site $k$ is favored for a nucleophilic attack, whereas if $\Delta\omega_k < 0$, then the site $k$ may be favored for an electrophilic attack. Because FFs are positive ($0 < f_k < 1$), $-1 < \Delta f_k < 1$, and the normalization condition for $\Delta\omega_k$ is

$$\sum_k \Delta\omega_k = \omega \sum_k \Delta f_k = 0 \tag{10}$$

Although $\Delta\omega_k$ and $\Delta f_k$ will contain the same intramolecular reactivity information the former is expected to be a better intermolecular descriptor because of its global information content.

We may analyze the nature of $\Delta\omega(\vec{r})$ in terms of that[22] of $\Delta f(\vec{r})$ as follows:

$$\left(\frac{\partial \omega(\vec{r})}{\partial N}\right)_v = \left(\frac{\partial\left[\omega f(\vec{r})\right]}{\partial N}\right)_v$$

$$= \left(\frac{\partial \omega}{\partial N}\right)_v f(\vec{r}) + \omega\left(\frac{\partial f(\vec{r})}{\partial N}\right)_v$$

$$= \left(\frac{\partial \omega}{\partial N}\right)_v f(\vec{r}) + \omega\,\Delta f(\vec{r})$$

$$= \left(\frac{\partial \omega}{\partial N}\right)_v f(\vec{r}) + \Delta\omega(\vec{r})$$

$$\Delta\omega(\vec{r}) = \left[\left(\frac{\partial \omega(\vec{r})}{\partial N}\right)_v - \left(\frac{\partial \omega}{\partial N}\right)_v f(\vec{r})\right]$$

The multiphilicity descriptor, $\Delta\omega(\vec{r})$ is a measure of the difference between local and global (modulated by $f(\vec{r})$) reactivity variations associated with the electron acceptance/ removal. Incidentally, the variation of $\left(\frac{\partial \omega}{\partial N}\right)$ across the periodic table is similar to that of μ.[23]

$$\left(\frac{\partial \omega}{\partial N}\right)_v = \frac{\partial}{\partial N}\left[\frac{\mu^2}{2\eta}\right]_v$$

$$= \frac{\mu}{\eta}\left(\frac{\partial \mu}{\partial \eta}\right)_v - \frac{\mu^2}{4\eta^2}\left(\frac{\partial \eta}{\partial N}\right)_v$$

$$= \frac{\mu}{\eta}\eta - \frac{\mu^2}{4\eta^2}\gamma$$

$$= \mu - \frac{\mu^2}{4\eta^2}\gamma = \mu - \frac{\omega\gamma}{2\eta}$$

Since $\gamma$ is generally very small,[24] $\left(\frac{\partial \omega}{\partial N}\right)_v$ is expected to follow the $\mu$ trend.

Problems associated with the definition of $\eta$ and the discontinuity[25] in E as a function of N will be present in the $\Delta f(\vec{r})$ definition and the discontinuity in $\rho(\vec{r})$.

Similar type of differentiation has also been attempted by other research workers.[26]

Also, to study the intra- and intermolecular reactivities another related descriptor namely, nucleophilicity excess ($\Delta\omega_g^{\mp}$) for a nucleophile, over the electrophilicity (net nucleophilicity) in it is defined as

$$\Delta\omega_g^{\mp} = \omega_g^{-} - \omega_g^{+} = \omega\left(f_g^{-} - f_g^{+}\right) \tag{11}$$

where $\omega_g^{-}(\equiv \sum_{k=1}^{n} \omega_k^{-})$ and $\omega_g^{+}(\equiv \sum_{k=1}^{n} \omega_k^{+})$ are the group philicities of the nucleophile in the molecule due to electrophilic and nucleophilic attacks respectively. It is expected that the nucleophilicity excess ($\Delta\omega_g^{\mp}$) for a nucleophile should always be positive whereas it will provide a negative value for an electrophile in a molecule.

In the present study, we use both the multiphilicity descriptor and nucleophilicity excess to probe the nature of attack/reactivity at a particular site in the selected systems.

## 3. Computational Details

The geometries of HCHO, $CH_3CHO$, $CH_3COCH_3$, $C_2H_5COC_2H_5$, $CH_2$=CHCHO $CH_3CH$=CHCHO, $NH_2OH$, $CH_3ONH_2$, $CH_3NHOH$, $OHCH_2CH_2NH_2$, $CH_3SNH_2$, $CH_3NHSH$, $SHCH_2CH_2NH_2$ and all-metal aromatic molecules, viz., $MAl_4^-$ (M=Li, Na, K and Cu) are optimized by B3LYP/6-311+G** as available in the *GAUSSIAN 98* package.[27] Various reactivity and selectivity descriptors such as chemical hardness, chemical potential, softness, electrophilicity and the appropriate local quantities employing natural population analysis (NPA)[28, 29] scheme are calculated. HPA scheme (Stockholder Partitioning Scheme) [30] as implemented in the $DMOL^3$ package [31] has also been used to calculate the local quantities employing BLYP/DND method. For all-metal aromatic molecules, ΔSCF method has been utilized to compute the ionization potential (IP) and electron affinity (EA) according to the equations ($I=E_{N-1} - E_N$, $A=E_N - E_{N+1}$, where $I$ and $A$ are obtained from total electronic energy calculations on the $N$-1, $N$, $N$+1-electron systems at the neutral molecule geometry).

## 4. Results and Discussion

A series of carbonyl compounds is selected in the present study to probe the usefulness of the multiphilicity descriptor (Figure 1). A comparison with various other descriptors and the recently derived dual descriptor is also probed. Due to bipolar nature of C=O bond, both nucleophilic and electrophilic attacks are possible at C and O sites. It is noted that the rate of nucleophilic addition on the carbonyl compound be reduced by electron donating alkyl groups and enhanced by electron withdrawing ones. [32] Recently, we have studied a set of these carbonyl compounds in the light of philicity and group philicity.[19] The global molecular properties of the selected series of carbonyl compounds are presented in Table 1. Various local quantities for particular sites of the selected systems are listed in Table 2 and Table 3. Selected compounds are grouped into two sets namely, nonconjugated and α, β-conjugated carbonyl compounds.

For the nonconjugated carbonyl compounds, the carbon atom ($C_1$) bearing the carbonyl group is expected to be the most reactive site towards a nucleophilic attack. Table 2 lists the values of local reactivity descriptors using B3LYP/6-311+G** method for NPA derived charges of the selected molecules. NPA derived local quantities predict the expected maximum value for carbonyl carbon ($C_1$) of all the selected molecules for $f_k^+$, $s_k^+$ and $\omega_k^+$. But $s_k^+/s_k^-$ is unable to provide the maximum value for $C_1$ atom due to negative FF values. One important point to note is that among the descriptors $f_k^+$, $s_k^+$, $\omega_k^+$ and $s_k^+/s_k^-$, $\omega_k^+$ value is capable of providing a clear distinction between carbonyl carbon ($C_1$) and the oxygen site for nucleophilic attack.

Since, HPA derived charges generally provide non-negative FF values, we also made use of it for local reactivity analysis on carbonyl compounds. HPA derived local reactivity descriptors also predict the expected maximum value for $C_1$ atom in the case of HCHO and $CH_3CHO$ but fails to predict for $CH_3COCH_3$ and $C_2H_5COC_2H_5$, where oxygen atom is shown to be prone towards nucleophilic attack. Nevertheless, the $f_k^+$ value of oxygen is almost same as that of carbonyl carbon ($C_1$), thus making it difficult to make a clear decision on the electrophilic behavior of these atoms. Under these situation, dual descriptors $\Delta f$ (r), $\Delta s_k$ and multiphilic descriptor $\Delta\omega$ (r), give a helping hand. All these quantities provide a clear difference between nucleophilic and electrophilic attacks at a particular site with their sign. That is, they provide positive value for site prone for nucleophilic attack and a negative value at the site prone for electrophilic attack. The advantage of multiphilic descriptor $\Delta\omega$ (r) is that they provide higher value in terms of magnitude compared to other dual descriptors. For instance, values of $\Delta f(r)$, $\Delta s_k$ and $\Delta\omega(r)$ for nucleophilic (electrophilic) attack at carbonyl carbon (oxygen) site of $CH_3CHO$ are 1.06 (-0.93), 0.17 (-0.15), 3.03 (-2.65) respectively for NPA derived charges. Almost the same trend is followed in the case of HPA derived charges.

The second group of compounds namely, α, β-conjugated carbonyl is elaborately studied in the recent past because of the presence of two reactive centers.[33] The first reactive site is the carbon ($C_1$) of the carbonyl, and the second is the carbon in the β position ($C_6$). In such a case, the β carbon is activated because of the withdrawing mesomeric effect of the adjacent carbonyl group. As seen from Table 2 and Table 3, NPA derived charges give a maximum value for $f_k^+$ to carbonyl carbon whereas HPA derived charges provide maximum $f_k^+$ value to the β carbon atom ($C_6$) in the case of $CH_2$=CHCHO molecule. For $CH_3$CH=CHCHO, NPA (HPA) provide maximum $f_k^+$ value of 0.44 (0.17) to carbonyl carbon ($C_1$) compared to the β carbon site of 0.34 (0.16). This ambiguous behavior may be due to the dependence of local reactivity descriptors on the selection of basis set and population schemes. Further oxygen site shows high value for $f_k^+$ and other local descriptors, making it difficult to predict the proper electrophilic site. Even now Δω (r) exhibits high positive value on both carbons that are supposed to be electrophilic and a high negative value on the oxygen site disclosing clearly its nucleophilic character compared to other dual descriptors. Also it can be noted from Tables 2 and 3 that, even for molecules with more than one reactive sites, Δω (r) is capable of making a clear distinction among them in terms of their magnitude. That is, for molecules 6 and 7 having two reactive sites as carbon ($C_1$) of the carbonyl and the carbon in the β position ($C_6$), our descriptors are capable of distinctly identifying the stronger site (electrophilic/nucleophilic).

Optimized structures along with atom numbering for the selected set of amines are presented in Figure 2. Global and local reactivity properties of the selected set of amines calculated using B3LYP/6-311+g** and BLYP/DND methods are presented in Tables 4 to 6. Global reactivity trend based on ω, is given by

*B3LYP/6-311+g** method* (Table 4)

(i) $CH_3ONH_2 > OHCH_2CH_2NH_2 > CH_3NHOH > NH_2OH$

(ii) $CH_3NHSH > SHCH_2CH_2NH_2 > CH_3SNH_2$

*BLYP/DND method* (Table 4)

(i) $CH_3ONH_2 > OHCH_2CH_2NH_2 > NH2OH > CH_3NHOH$

(ii) $CH_3NHSH > SHCH_2CH_2NH_2 > CH_3SNH_2$

Though both the methods show variation in reactivity trend for oxygen containing systems, trends related to sulfur containing systems are same.

Based on NPA and HPA charge derived multiphilic descriptor at nitrogen site ($\Delta\omega_N$), following reactivity trend has been obtained,

*NPA* (Table 5)

(1) $OHCH_2CH_2NH_2 > CH_3NHOH > NH_2OH > CH_3ONH_2$

(2) $CH_3NHSH > SHCH_2CH_2NH_2 > CH_3SNH_2$

*HPA* (Table 6)

(1) $OHCH_2CH_2NH_2 > CH_3ONH_2 > NH_2OH > CH_3NHOH$

(2) $CH_3NHSH > SHCH_2CH_2NH_2 > CH_3SNH_2$

It may be noted that trends are same as ω for sulfur containing systems, but shows variations with respect to oxygen containing systems for both NPA and HPA charge derived $\Delta\omega_N$.

So for as the intramolecular reactivity trends are concerned, site with maximum negative value of $\Delta\omega_k$ is the most preferred site for electrophilic attack. Chemical intuition suggests that N site is more prone towards electrophilic attack. Table 7 lists the site with maximum negative value for $\Delta\omega_k$ for the selected set of amines. It is seen that with a few exception, N site is predicted as the most preferred site for electrophilic attack.

Further in order to test $\Delta\omega_k$ along intrinsic reaction coordinate (IRC), we consider a cope rearrangement of hexa-1,5-diene. This is an example of [3,3] sigmatropic reaction. Figure 3 provides the optimized geometrical structures with atom numbering for the reactant, transition state and product calculated using B3LYP/6-31G* level of theory. Table 8 gives the global reactivity parameters of the reactant, transition state and product. As expected, hardness is minimum (2.48 eV) and the corresponding electrophilicity index is maximum (1.57 eV) at the transition state. Variation of global reactivity parameter along the IRC path is presented in Table 9 and Figure 4 (a-b). Variation of energy (E) and ω along IRC path is given in Figure 5a. It is seen that both E and ω are maximum around the transition state indicating it as the most unstable structure along the IRC path. Figure 5 b provides the variation of hardness ($\eta$) and polarizability ($\alpha$) along the IRC path. An inverse relationship exists between them. That is, $\eta$ reaches a minimum whereas $\alpha$ becomes maximum at the transition state as expected.

Variation of multiphilic descriptor ($\Delta\omega_k$) along IRC for the important atomic sites ($C_1$ and $C_3$/ $C_6$ and $C_{11}$) is presented in Figure 5. In going from reactant to product, $C_1$ and $C_3$ ($C_6$ and $C_{11}$) sites change their nature and become more prone towards electrophilic attack (nucleophilic attack) at the product side. This change in the nature of attack takes place around the transition state.

In studying the importance of nucleophilicity excess ($\Delta\omega_g^{\mp}$) descriptor, a careful analysis on the electronic structure, property and reactivity of all-metal aromatic compounds, viz., $MAl_4^-$ (M=Li, Na, K and Cu) is performed. The four membered aluminum unit $Al_4$ present in all the molecules may be considered as a single unit. This unit can easily take part in charge transfer process with the M (≡Li, Na, K, Cu) atom in those complexes.

Figure 6 shows the various stable isomers of $MAl_4^-$. The $C_{4v}$ isomer of the $MAl_4^-$ is reported as energetically most stable, least polarizable and hardest.[34, 35] Table 10 presents the group philicity ($\omega_g^+$, $\omega_g^-$) values of the $Al_4^{2-}$ nucleophile and $M^+$ (M=Li, Na, K, Cu) electrophile in the $MAl_4^-$ isomers. It is found that in all $MAl_4^-$ isomers the nucleophilicity of the $Al_4^{2-}$ aromatic unit overwhelms its electrophilic trend (i.e. $\omega_g^- \succ \omega_g^+$) and therefore $\Delta\omega_g^{\mp}$ is positive, whereas the electrophilicity of $M^+$ dominates over its nucleophilicity (i.e. $\omega_g^+ \succ \omega_g^-$) and therefore $\Delta\omega_g^{\mp}$ is negative as expected. It is important to note that $\Delta\omega_g^{\mp}$ of $Al_4^{2-}$ is maximum in the case of most stable $C_{4v}$ isomer of the $MAl_4^-$ molecule. The order of the $\Delta\omega_g^{\mp}$ value of $Al_4^{2-}$ nucleophile in $MAl_4^-$, $C_{4v} \succ C_{2v} \succ C_{\infty v}$, i.e. stabilization of an $MAl_4^-$ isomer (except in $KAl_4^-$) increases its nucleophilicity and accordingly can be used as a better molecular cathode. It is also important to note that the nucleophilicity of the $Al_4^{2-}$ unit in $MAl_4^-$ ($C_{4v}$) increases as $K \prec Cu \prec Na \prec Li$ according to the respective nucleophilicity excess values. Standard expressions[1-5] for ΔN and ΔE in terms of group electronegativity and group hardness will provide additional insights into the electron transfer process.

Variation of $\Delta\omega_k$ along the IRC of three selected reactions,[36] viz., a) a thermoneutral reaction: $F_a^- + CH_3\text{-}F_b \rightarrow F_a\text{-}CH_3 + F_b^-$, b) an endothermic reaction: HNO → HON, c) an exothermic reaction: $H_2OO$ → HOOH is provided in figures 7 (a) – 7(c). For the thermoneutral reaction, both the $F_a^-$ (bond making) and $F_b^-$ (bond breaking) are nucleophilic. The net nucleophilicity of the $F_a^-$ atom is more than that of the $F_b^-$ atom along the IRC from reactant side to TS and the situation is reversed for the IRCs pertaining to the TS to product side. For the endothermic reaction, the net nucleophilicity of O (bond making) is higher than that of N (bond breaking) along the IRC. In the case of exothermic reaction, the O1 (bond making) atom is more electrophilic than its nucleophilic activity. Moreover, its Fukui function values as calculated through Mulliken Population Analysis (MPA) scheme become negative in some cases. For the thermoneutral reaction $\Delta\omega_k$ is minimum at the transition state. For other two reactions, $\Delta\omega_k$ does not always follow the trend that the IRC corresponding to the minimum value of $\omega_k^{\pm}$ (if not zero) is in accordance with the Hammond's postulate.[36] Figures 8 (a) – 8 (c) provide the profiles for the corresponding reaction forces.[37]

Apart from the important points corresponding to  the reactant (R), the transition state (TS) and the product (P) there exists two other important points associated with the configurations having the force maximum ($F_{max}$) and the force minimum ($F_{min}$). The zeroes, maxima and minima of the reaction force define key points along the reaction coordinate, which divide it into three reaction regions that are identified through vertical dashed lined in Figure 8. The first stage, in the reactant region, tends to be preparative in nature with emphasis in structural effects such as rotation, bond stretching, angle bending, etc., that will facilitate subsequent steps. The transition state region is mostly characterized by electronic rearrangements whereas the product region is mainly associated to structural relaxation necessary to reach the products. We have shown that analyzing a chemical reaction in terms of these regions can provide significant insight into its mechanism and the roles played by external factors, such as external potentials and solvents.[37, 38] Partition of the activation energies in terms of the work done in going from i) R to $F_{min}$: $W_1$, ii) $F_{min}$ to TS: $W_2$, iii) TS to $F_{max}$: $W_3$ and iv) $F_{max}$ to P: $W_4$ gives the activation energy for the forward reaction ($E_f^{\#}$) as ($W_1+W_2$) and that of the reverse reaction ($E_r^{\#}$) as $-(W_3+W_4)$. Therefore the reaction energy ($\Delta E^0$) becomes ($E_f^{\#} - E_r^{\#} = W_1+W_2+W_3+W_4$). These values are provided in Table 11. As expected $\Delta E^0$ is zero, negative and positive for the thermoneutral, exothermic and endothermic reactions respectively. The skew-symmetric nature of the force profile for the thermoneutral reaction suggests that $A=W_1+W_4$ and $B=W_2+W_3$ would be zero. Similarly A, B would be positive (negative) for the endo(exo)thermic reactions. The transition state at the IRC=0 configuration lies at the middle between $F_{max}$ and $F_{min}$ configurations for the thermoneutral reaction whereas it lies towards the $F_{min}(F_{max})$ configurations for the exo(endo)thermic reaction, a signature of the Hammond postulate via reaction force.

Similar values of $W_1$ and $W_2$ (see Table 11) together with the changes observed in the nucleophilicity along the reaction coordinate for the thermoneutral $S_N2$ substitution and for the exothermic reaction $H_2OO \rightarrow HOOH$ indicate that structural and electronic reordering show up at the very beginning of the reaction, [37,38] through a sharp decrease of the nucleophilicity, this change practically ceases at the transition state of the exothermic reaction to reach the product value. It is interesting to note that in both cases the lowering of nucleophilicity of the key atoms from the reactants ($\Delta\omega(F_a/F_b) \sim 0.014$; $\Delta\omega(O1) \sim 0.14$) to the transition state ($\Delta\omega(F_a/F_b) \sim 0.004$; $\Delta\omega(O1) \sim 0.0$) requires a similar amount of energy (9.54 kcal/mol and 7.39 kcal/mol, respectively). It can be observed in Table 11 that for the thermoneutral reaction $W_1 > W_2$ indicating that the preparation step requires more energy than the transition to product step. On the other hand, the $W_2$ values for the thermoneutral and exothermic reactions are quite close to each other and the work $W_1$ associated to the preparation step in the thermoneutral reaction is larger than that of the exothermic reaction, this indicates that in the $S_N2$ reaction the structural reordering of the $CH_3$ group to reach the $D_{3h}$ structure at the transition state is the key transformation that involve most of the activation energy. In the endothermic HNO $\rightarrow$ HON reaction the small changes of nucleophilicity together with large values of $W_1$ and $W_2$ indicates that the reaction is mainly driven by the structural reordering in the preparation step.

## 5. Conclusions

A multiphilicity descriptor ($\Delta\omega_k$) is proposed and tested in this work. It is shown that, $\Delta\omega_k$ helps in identifying the electrophilic/nucleophilic nature of a specific site within a molecule. A comparison between different local reactivity descriptors is carried out on a set of carbonyl compounds. Also a selected set of amines is analyzed using $\Delta\omega_k$. Further, we also consider a cope rearrangement of hexa-1,5-diene to test the variation of $\Delta\omega_k$ along IRC path. It is seen that $\Delta\omega_k$ presents a clear distinction between electrophilic and nucleophilic sites within a molecule in terms of their magnitude and sign. Hence they reveal the fact that multiphilic descriptor can effectively be used in characterizing the electrophilic/nucleophilic nature of a given site in a molecule. Also the importance of nucleophilicity excess ($\Delta\omega_g^{\mp}$) descriptor on the reactivity of all-metal aromatic compounds, viz., $MAl_4^-$ (M=Li, Na, K and Cu) is successfully analyzed. Important insight into three different types of reactions, viz., a) thermoneutral, b) endothermic and c) exothermic are obtained through the analysis of the multiphilic descriptor profiles within the reaction regions defined by reaction force along the reaction path.

The results discussed so far clearly show the importance of the selected descriptors, namely, multiphilic descriptor and nucleophilicity excess in analyzing the overall reactivity trends in molecular systems.

**Acknowledgment:**

PKC and DRR thank BRNS, Mumbai for financial assistance. JP and BSK thank the IIT Kharagpur for providing the facilities required for a summer project. JP also thanks the UGC for selecting him to carryout his Ph.D. work under FIP. ATL and SGO wish to thank financial support from FONDECYT, grant N° 1060590, FONDAP through project N° 11980002 (CIMAT) and Programa Bicentenario en Ciencia y Tecnología (PBCT), Proyecto de Inserción Académica N° 8. ATL is also indebted to the John Simon Guggenheim Foundation for a fellowship.

**TABLE 1: Calculated Global Reactivity Properties of the Selected Molecules using B3LYP/6-311+g** and BLYP/DND method.**

| | η | μ | ω | S | η | μ | ω | S |
|---|---|---|---|---|---|---|---|---|
| Molecules | B3LYP/6-311+g** (eV) | | | | BLYP/DND (eV) | | | |
| HCHO | 2.960 | -4.707 | 3.742 | 0.169 | 1.942 | -4.260 | 4.673 | 0.258 |
| $CH_3CHO$ | 3.115 | -4.224 | 2.864 | 0.161 | 2.096 | -3.791 | 3.425 | 0.238 |
| $CH_3COCH_3$ | 3.144 | -3.910 | 2.432 | 0.159 | 2.133 | -3.456 | 2.800 | 0.234 |
| $C_2H_5COC_2H_5$ | 3.153 | -3.799 | 2.288 | 0.159 | 2.151 | -3.367 | 2.635 | 0.233 |
| $CH_2$=CHCHO | 2.503 | -4.904 | 4.805 | 0.200 | 1.545 | -4.413 | 6.303 | 0.324 |
| $CH_3$CH=CHCHO | 2.542 | -4.631 | 4.217 | 0.197 | 1.593 | -4.132 | 5.359 | 0.314 |

**TABLE 2: Calculated Local Reactivity Properties of the Selected Molecules using B3LYP/6-311+g** method for NPA derived charges.**

| Molecule | | $f_k^+$ | $f_k^-$ | $s_k^+$ | $s_k^-$ | $s_k^+/s_k^-$ | $\omega_k^+$ | $\omega_k^-$ | $\Delta f_k$ $=f_k^+- f_k^-$ | $\Delta s_k$ $=s_k^+-s_k^-$ | $\Delta\omega_k$ $=\omega_k^+-\omega_k^-$ |
|---|---|---|---|---|---|---|---|---|---|---|---|
| HCHO | C | 0.8323 | -0.1722 | 0.1406 | -0.0291 | -4.8331 | 3.1146 | -0.6444 | 1.0045 | 0.1697 | 3.7591 |
| | O | 0.0399 | 0.9409 | 0.0067 | 0.1589 | 0.0424 | 0.1494 | 3.5211 | -0.9010 | -0.1522 | -3.3718 |
| $CH_3CHO$ | C1 | 0.8178 | -0.2416 | 0.1313 | -0.0388 | -3.3856 | 2.3419 | -0.6917 | 1.0593 | 0.1700 | 3.0337 |
| | O | 0.0072 | 0.9320 | 0.0012 | 0.1496 | 0.0077 | 0.0206 | 2.6691 | -0.9250 | -0.1484 | -2.6485 |
| $CH_3COCH_3$ | C1 | 0.3142 | -0.2916 | 0.0500 | -0.0464 | -1.0772 | 0.7640 | -0.7092 | 0.6058 | 0.0964 | 1.4732 |
| | O | -0.2540 | 0.9286 | -0.0404 | 0.1477 | -0.2734 | -0.6170 | 2.2582 | -1.1820 | -0.1881 | -2.8755 |
| $C_2H_5COC_2H_5$ | C1 | 0.3064 | -0.2944 | 0.0486 | -0.0467 | -1.0408 | 0.7011 | -0.6736 | 0.6007 | 0.0953 | 1.3746 |
| | O | -0.2650 | 0.8751 | -0.0420 | 0.1388 | -0.3024 | -0.606 | 2.0025 | -1.1400 | -0.1807 | -2.6080 |
| $CH_2$=CHCHO | C6 | 0.2789 | 0.2070 | 0.0557 | 0.0413 | 1.3472 | 1.3402 | 0.9944 | 0.0719 | 0.0144 | 0.3458 |
| | C1 | 0.4355 | -0.2288 | 0.0870 | -0.0457 | -1.9033 | 2.0926 | -1.0995 | 0.6643 | 0.1327 | 3.1921 |
| | O | -0.0560 | 0.9265 | -0.0112 | 0.1851 | -0.0605 | -0.2700 | 4.4518 | -0.9830 | -0.1963 | -4.7213 |
| $CH_3$CH=CHCHO | C6 | 0.3437 | 0.0926 | 0.0676 | 0.0182 | 3.7143 | 1.4494 | 0.3904 | 0.2511 | 0.0494 | 1.0590 |
| | C1 | 0.4408 | -0.2365 | 0.0867 | -0.0465 | -1.8642 | 1.8592 | -0.9973 | 0.6773 | 0.1332 | 2.8566 |
| | O | -0.0670 | 0.9281 | -0.0132 | 0.1825 | -0.0721 | -0.2820 | 3.9142 | -0.9950 | -0.1957 | -4.1964 |

**TABLE 3: Calculated Local Reactivity Properties of the Selected Molecules using BLYP/DND method for HPA derived charges.**

| Molecule | | $f_k^+$ | $f_k^-$ | $s_k^+$ | $s_k^-$ | $s_k^+/s_k^-$ | $\omega_k^+$ | $\omega_k^-$ | $\Delta f_k = f_k^+ - f_k^-$ | $\Delta s_k = s_k^+ - s_k^-$ | $\Delta\omega_k = \omega_k^+ - \omega_k^-$ |
|---|---|---|---|---|---|---|---|---|---|---|---|
| HCHO | C | 0.3973 | 0.2373 | 0.1023 | 0.0611 | 1.6744 | 1.8563 | 1.1088 | 0.1600 | 0.0412 | 0.7476 |
| | O | 0.3010 | 0.4232 | 0.0775 | 0.1090 | 0.7113 | 1.4064 | 1.9774 | -0.1222 | -0.0315 | -0.5710 |
| $CH_3CHO$ | C1 | 0.2998 | 0.1642 | 0.0715 | 0.0391 | 1.8267 | 1.0268 | 0.5624 | 0.1356 | 0.0324 | 0.4644 |
| | O | 0.2708 | 0.3782 | 0.0646 | 0.0902 | 0.7165 | 0.9275 | 1.2953 | -0.1074 | -0.0256 | -0.3678 |
| $CH_3COCH_3$ | C1 | 0.2108 | 0.1154 | 0.0494 | 0.0271 | 1.8262 | 0.5902 | 0.3231 | 0.0954 | 0.0223 | 0.2671 |
| | O | 0.2359 | 0.3499 | 0.0553 | 0.0820 | 0.6742 | 0.6605 | 0.9797 | -0.1140 | -0.0267 | -0.3192 |
| $C_2H_5COC_2H_5$ | C1 | 0.1346 | 0.0990 | 0.0313 | 0.0230 | 1.3598 | 0.3547 | 0.2609 | 0.0356 | 0.0083 | 0.0938 |
| | O | 0.1449 | 0.2873 | 0.0337 | 0.0668 | 0.5045 | 0.3818 | 0.7570 | -0.1424 | -0.0331 | -0.3752 |
| $CH_2$=CHCHO | C1 | 0.1780 | 0.1357 | 0.0577 | 0.0440 | 1.3117 | 1.1219 | 0.8553 | 0.0423 | 0.0137 | 0.2666 |
| | C6 | 0.2062 | 0.1253 | 0.0668 | 0.0406 | 1.6457 | 1.2997 | 0.7898 | 0.0809 | 0.0262 | 0.5099 |
| | O | 0.1797 | 0.3414 | 0.0582 | 0.1106 | 0.5264 | 1.1326 | 2.1518 | -0.1620 | -0.0524 | -1.0191 |
| $CH_3CH$=CHCHO | C6 | 0.1592 | 0.1114 | 0.0500 | 0.0350 | 1.4291 | 0.8532 | 0.5970 | 0.0478 | 0.0150 | 0.2562 |
| | C1 | 0.1741 | 0.1095 | 0.0547 | 0.0344 | 1.5900 | 0.9330 | 0.5868 | 0.0646 | 0.0203 | 0.3462 |
| | O | 0.1739 | 0.2450 | 0.0546 | 0.0769 | 0.7098 | 0.9319 | 1.3130 | -0.0710 | -0.0223 | -0.3810 |

**TABLE 4: Calculated Global Reactivity Properties of the Selected Molecules using B3LYP/6-311+g** and BLYP/DND method.**

| | $\eta$ | $\mu$ | $\omega$ | S | $\eta$ | $\mu$ | $\omega$ | S |
|---|---|---|---|---|---|---|---|---|
| Molecules | B3LYP/6-311+g** (eV) | | | | BLYP/DND (eV) | | | |
| $NH_2OH$ | 3.869 | -3.553 | 1.632 | 0.129 | 3.411 | -1.399 | 0.287 | 0.147 |
| $CH_3ONH_2$ | 3.630 | -3.738 | 1.925 | 0.138 | 3.549 | -3.053 | 1.313 | 0.141 |
| $CH_3NHOH$ | 3.482 | -3.392 | 1.652 | 0.144 | 3.229 | -1.308 | 0.265 | 0.155 |
| $OHCH_2CH_2NH_2$ | 3.343 | -3.507 | 1.840 | 0.150 | 3.348 | -2.689 | 1.080 | 0.149 |
| $CH_3SNH_2$ | 3.050 | -3.331 | 1.819 | 0.164 | 2.447 | -1.750 | 0.626 | 0.204 |
| $CH_3NHSH$ | 3.148 | -3.629 | 2.092 | 0.159 | 2.466 | -3.596 | 2.622 | 0.203 |
| $SHCH_2CH_2NH_2$ | 3.135 | -3.417 | 1.862 | 0.159 | 2.521 | -1.843 | 0.674 | 0.198 |

**TABLE 5: Calculated Local Reactivity Properties of the Selected Molecules using B3LYP/6-311+g** method for NPA derived charges.**

| Molecule | | $f_k^+$ | $f_k^-$ | $s_k^+$ | $s_k^-$ | $s_k^-/s_k^+$ | $\omega_k^+$ | $\omega_k^-$ | $\Delta f_k = f_k^+ - f_k^-$ | $\Delta s_k = s_k^+ - s_k^-$ | $\Delta\omega_k = \omega_k^+ - \omega_k^-$ |
|---|---|---|---|---|---|---|---|---|---|---|---|
| $NH_2OH$ | N | 0.1870 | 0.4140 | 0.0274 | 0.0607 | 2.2139 | 0.0536 | 0.1187 | -0.2270 | -0.0333 | -0.0651 |
| | O | 0.2390 | 0.2300 | 0.0350 | 0.0337 | 0.9623 | 0.0685 | 0.0659 | 0.0090 | 0.0013 | 0.0026 |
| $CH_3ONH_2$ | C | 0.0870 | 0.0680 | 0.1410 | 1.3130 | 0.0123 | 0.0096 | 0.7816 | 0.1142 | 0.0893 | 0.0190 |
| | N | 0.1500 | 0.3510 | 0.0211 | 0.0495 | 2.3400 | 0.1969 | 0.4608 | -0.2010 | -0.0283 | -0.2639 |
| | O | 0.0720 | 0.1740 | 0.0101 | 0.0245 | 2.4167 | 0.0945 | 0.2284 | -0.1020 | -0.0144 | -0.1339 |
| $CH_3NHOH$ | C | 0.0470 | 0.0740 | 0.0073 | 0.0115 | 1.5745 | 0.0124 | 0.0196 | -0.0270 | -0.0042 | -0.0071 |
| | N | 0.1200 | 0.3390 | 0.0186 | 0.0525 | 2.8250 | 0.0318 | 0.0898 | -0.2190 | -0.0339 | -0.0580 |
| | O | 0.2100 | 0.1770 | 0.0325 | 0.0274 | 0.8429 | 0.0556 | 0.0469 | 0.0330 | 0.0051 | 0.0087 |
| $OHCH_2CH_2NH_2$ | $C_1$ | 0.0540 | 0.0330 | 0.0081 | 0.0049 | 0.6111 | 0.0583 | 0.0356 | 0.0210 | 0.0031 | 0.0227 |
| | $C_2$ | 0.0400 | 0.0610 | 0.006 | 0.0091 | 1.5250 | 0.0432 | 0.0659 | -0.0210 | -0.0031 | -0.0227 |
| | N | 0.0630 | 0.3470 | 0.0094 | 0.0518 | 5.5079 | 0.0680 | 0.3746 | -0.2840 | -0.0424 | -0.3066 |
| | O | 0.1400 | 0.1010 | 0.0209 | 0.0151 | 0.7214 | 0.1511 | 0.1090 | 0.0390 | 0.0058 | 0.0421 |
| $CH_3SNH_2$ | C | 0.0550 | 0.0640 | 0.0112 | 0.0131 | 1.1636 | 0.0344 | 0.0400 | -0.0090 | -0.0018 | -0.0056 |
| | N | 0.1490 | 0.0820 | 0.0305 | 0.0168 | 0.5503 | 0.0932 | 0.0513 | 0.0670 | 0.0137 | 0.0419 |
| | S | 0.3580 | 0.5510 | 0.0732 | 0.1126 | 1.5391 | 0.2239 | 0.3447 | -0.1930 | -0.0394 | -0.1207 |
| $CH_3NHSH$ | C | 0.0530 | 0.0540 | 0.0107 | 0.0110 | 1.0189 | 0.1390 | 0.1416 | -0.0010 | -0.0002 | -0.0026 |
| | N | 0.1310 | 0.1740 | 0.0266 | 0.0353 | 1.3282 | 0.3434 | 0.4562 | -0.0430 | -0.0087 | -0.1127 |
| | S | 0.4530 | 0.4420 | 0.0919 | 0.0896 | 0.9757 | 1.1876 | 1.1588 | 0.0110 | 0.0022 | 0.0288 |
| $SHCH_2CH_2NH_2$ | $C_1$ | 0.0780 | 0.0410 | 0.0155 | 0.0081 | 0.5256 | 0.0525 | 0.0276 | 0.0370 | 0.0073 | 0.0249 |
| | $C_2$ | 0.0290 | 0.0250 | 0.0058 | 0.0050 | 0.8621 | 0.0195 | 0.0168 | 0.0040 | 0.0008 | 0.0027 |
| | N | 0.0380 | 0.1270 | 0.0075 | 0.0252 | 3.3421 | 0.0256 | 0.0856 | -0.0890 | -0.0177 | -0.0600 |
| | S | 0.3890 | 0.4710 | 0.0772 | 0.0934 | 1.2108 | 0.2621 | 0.3173 | -0.0820 | -0.0163 | -0.0552 |

**TABLE 6 Calculated Local Reactivity Properties of the Selected Molecules using BLYP/DND method for HPA derived charges.**

| Molecule | | $f_k^+$ | $f_k^-$ | $s_k^+$ | $s_k^-$ | $s_k^-/s_k^+$ | $\omega_k^+$ | $\omega_k^-$ | $\Delta f_k = f_k^+ - f_k^-$ | $\Delta s_k = s_k^+ - s_k^-$ | $\Delta\omega_k = \omega_k^+ - \omega_k^-$ |
|---|---|---|---|---|---|---|---|---|---|---|---|
| $NH_2OH$ | N | 0.1837 | 0.9327 | 0.0237 | 0.1205 | 5.0777 | 0.2997 | 1.5218 | -0.7490 | -0.0970 | -1.2220 |
| | O | -0.0770 | 0.5114 | -0.0100 | 0.0661 | -6.6170 | -0.1261 | 0.8344 | -0.5890 | -0.0760 | -0.9610 |
| $CH_3ONH_2$ | C | 0.5410 | 0.0819 | 0.0746 | 0.0113 | 0.1513 | 1.0412 | 0.1576 | 0.4592 | 0.0633 | 0.8837 |
| | N | -0.1510 | 0.2534 | -0.0210 | 0.0349 | -1.6740 | -0.2913 | 0.4877 | -0.4050 | -0.0560 | -0.7790 |
| | O | -0.1790 | 0.9011 | -0.0250 | 0.1242 | -5.0267 | -0.3450 | 1.7342 | -1.0800 | -0.1490 | -2.0790 |
| $CH_3NHOH$ | C | 0.4598 | 0.1677 | 0.0660 | 0.0241 | 0.3647 | 0.7598 | 0.2771 | 0.2921 | 0.0419 | 0.4827 |
| | N | -0.0580 | 0.7950 | -0.0080 | 0.1142 | -13.725 | -0.0957 | 1.3136 | -0.8530 | -0.1220 | -1.4090 |
| | O | -0.2690 | 0.4537 | -0.0390 | 0.0651 | -1.6855 | -0.4448 | 0.7497 | -0.7230 | -0.1040 | -1.1940 |
| $OHCH_2CH_2NH_2$ | $C_1$ | 0.1186 | 0.0254 | 0.0177 | 0.0038 | 0.2140 | 0.2181 | 0.0467 | 0.0932 | 0.0139 | 0.1715 |
| | $C_2$ | 0.4003 | 0.1067 | 0.0599 | 0.0160 | 0.2666 | 0.7365 | 0.1964 | 0.2936 | 0.0439 | 0.5401 |
| | N | -0.3040 | 0.9520 | -0.0450 | 0.1424 | -3.1337 | -0.5589 | 1.7514 | -1.2560 | -0.1880 | -2.3100 |
| | O | -0.3340 | 0.5965 | -0.0500 | 0.0892 | -1.7842 | -0.6151 | 1.0974 | -0.9310 | -0.1390 | -1.7120 |
| $CH_3SNH_2$ | C | 0.0667 | 0.3358 | 0.0100 | 0.0502 | 5.0377 | 0.1226 | 0.6178 | -0.2690 | -0.0400 | -0.4950 |
| | N | -0.297 | 0.4790 | -0.044 | 0.0717 | -1.6119 | -0.5467 | 0.8813 | -0.7760 | -0.1160 | -1.4280 |
| | S | 0.3671 | 0.6485 | 0.0549 | 0.0970 | 1.7667 | 0.6753 | 1.1931 | -0.2810 | -0.0420 | -0.5180 |
| $CH_3NHSH$ | C | 0.1715 | 0.1732 | 0.0256 | 0.0259 | 1.0100 | 0.3154 | 0.3186 | -0.0020 | -0.0003 | -0.0030 |
| | N | -0.225 | 0.9064 | -0.0340 | 0.1356 | -4.0267 | -0.4141 | 1.6676 | -1.1320 | -0.1690 | -2.0820 |
| | S | 0.3479 | 0.2249 | 0.0520 | 0.0336 | 0.6465 | 0.6400 | 0.4137 | 0.1230 | 0.01840 | 0.2262 |
| $SHCH_2CH_2NH_2$ | $C_1$ | 0.0117 | 0.2268 | 0.0017 | 0.0339 | 19.432 | 0.0215 | 0.4172 | -0.2150 | -0.0320 | -0.3960 |
| | $C_2$ | 0.1651 | 0.0876 | 0.0247 | 0.0131 | 0.5309 | 0.3037 | 0.1612 | 0.0774 | 0.0116 | 0.1425 |
| | N | -0.292 | 0.7628 | -0.0440 | 0.1141 | -2.6164 | -0.5364 | 1.4035 | -1.0540 | -0.1580 | -1.9400 |
| | S | 0.1064 | 0.5646 | 0.0159 | 0.0845 | 5.3089 | 0.1957 | 1.0388 | -0.4580 | -0.0690 | -0.8430 |

**TABLE 7: Atomic site with maximum value for multiphilic descriptor ($\Delta\omega_k$) for the selected set of amines.**

| molecule | site with maximum value for $\Delta\omega_k$ | |
|---|---|---|
| | NPA | HPA |
| $NH_2OH$ | N | N |
| $CH_3ONH_2$ | O | N |
| $CH_3NHOH$ | N | N |
| $OHCH_2CH_2NH_2$ | N | N |
| $CH_3SNH_2$ | N | S |
| $CH_3NHSH$ | N | N |
| $SHCH_2CH_2NH_2$ | N | N |

**TABLE 8: Global reactivity descriptors calculated at B3LYP/6-31G* level of theory.**

| Species | $\eta$ (eV) | $\mu$ (eV) | $\omega$ (eV) |
|---|---|---|---|
| Reactant | 3.64 | -2.89 | 1.15 |
| Transition State | 2.48 | -2.79 | 1.57 |
| Product | 3.64 | -2.89 | 1.15 |

**TABLE 9**: **Global reactivity descriptors along the intrinsic reaction coordinate calculated at B3LYP/6-31G* level of theory.**

| Points along IRC | *E* (Hartrees) | *η* (eV) | *μ* (eV) | *ω* (eV) | *α* (a.u.) |
|---|---|---|---|---|---|
| 1 | -234.5673091 | 2.65 | -2.7825 | 1.46 | 64.94 |
| 2 | -234.5661087 | 2.63 | -2.7827 | 1.47 | 65.21 |
| 3 | -234.5649450 | 2.61 | -2.7828 | 1.49 | 65.47 |
| 4 | -234.5638273 | 2.59 | -2.7836 | 1.50 | 65.74 |
| 5 | -234.5627655 | 2.57 | -2.7836 | 1.51 | 65.98 |
| 6 | -234.5617681 | 2.55 | -2.7843 | 1.52 | 66.22 |
| 7 | -234.5608445 | 2.54 | -2.7843 | 1.53 | 66.42 |
| 8 | -234.5600030 | 2.53 | -2.7851 | 1.54 | 66.63 |
| 9 | -234.5592516 | 2.51 | -2.7852 | 1.54 | 66.80 |
| 10 | -234.5585980 | 2.50 | -2.7859 | 1.55 | 66.96 |
| 11 | -234.5580104 | 2.50 | -2.7857 | 1.56 | 67.07 |
| 12 | -234.5575677 | 2.49 | -2.7866 | 1.56 | 67.20 |
| 13 | -234.5575677 | 2.49 | -2.7866 | 1.56 | 67.20 |
| 14 | -234.5580104 | 2.50 | -2.7857 | 1.56 | 67.07 |
| 15 | -234.5585980 | 2.50 | -2.7859 | 1.55 | 66.96 |
| 16 | -234.5592516 | 2.51 | -2.7852 | 1.54 | 66.80 |
| 17 | -234.5600030 | 2.53 | -2.7851 | 1.54 | 66.63 |
| 18 | -234.5608445 | 2.54 | -2.7843 | 1.53 | 66.42 |
| 19 | -234.5617681 | 2.55 | -2.7843 | 1.52 | 66.22 |
| 20 | -234.5627655 | 2.57 | -2.7836 | 1.51 | 65.98 |
| 21 | -234.5638273 | 2.59 | -2.7836 | 1.50 | 65.74 |
| 22 | -234.5649450 | 2.61 | -2.7830 | 1.49 | 65.47 |
| 23 | -234.5661087 | 2.63 | -2.7827 | 1.47 | 65.21 |
| 24 | -234.5673092 | 2.65 | -2.7825 | 1.46 | 64.94 |

**TABLE 10: Group Philicity ($\omega_g^+, \omega_g^-$) Values for Nucleophilic and Electrophilic Attacks Respectively for the Ionic Units of Different Isomers of $LiAl_4^-$, $NaAl_4^-$, $KAl_4^-$ and $CuAl_4^-$.**

| Isomers | Ionic Unit | $\omega_g^+$ | $\omega_g^-$ | $\Delta\omega_g^{\mp}$ |
|---|---|---|---|---|
| $LiAl_4^-$ | $Al_4^{2-}$ | 0.0070 | 0.0095 | 0.0025 |
| ($C_{\infty v}$) | $Li^+$ | 0.0063 | 0.0037 | -0.0025 |
| $LiAl_4^-$ | $Al_4^{2-}$ | 1.3E-05 | 0.0055 | 0.0055 |
| ($C_{2v}$) | $Li^+$ | 0.0068 | 0.0013 | -0.0055 |
| $LiAl_4^-$ | $Al_4^{2-}$ | -0.0372 | 0.2965 | 0.3338 |
| ($C_{4v}$) | $Li^+$ | 0.4055 | 0.0718 | -0.3338 |
| $NaAl_4^-$ | $Al_4^{2-}$ | 0.0070 | 0.0102 | 0.0032 |
| ($C_{\infty v}$) | $Na^+$ | 0.0074 | 0.0042 | -0.0032 |
| $NaAl_4^-$ | $Al_4^{2-}$ | -0.0001 | 0.0078 | 0.0079 |
| ($C_{2v}$) | $Na^+$ | 0.0096 | 0.0017 | -0.0079 |
| $NaAl_4^-$ | $Al_4^{2-}$ | -0.0073 | 0.1024 | 0.1097 |
| ($C_{4v}$) | $Na^+$ | 0.1301 | 0.0204 | -0.1097 |
| $KAl_4^-$ | $Al_4^{2-}$ | 0.0044 | 0.0095 | 0.0051 |
| ($C_{\infty v}$) | $K^+$ | 0.0106 | 0.0054 | -0.0051 |
| $KAl_4^-$ | $Al_4^{2-}$ | 0.0023 | 0.0101 | 0.0078 |
| ($C_{2v}$) | $K^+$ | 0.0118 | 0.0039 | -0.0078 |
| $KAl_4^-$ | $Al_4^{2-}$ | 0.0008 | 0.0066 | 0.0057 |
| ($C_{4v}$) | $K^+$ | 0.0078 | 0.0021 | -0.0057 |
| $CuAl_4^-$ | $Al_4^{2-}$ | 0.0031 | 0.0036 | 0.0006 |
| ($C_{\infty v}$) | $Cu^+$ | 0.0014 | 0.0009 | -0.0006 |
| $CuAl_4^-$ | $Al_4^{2-}$ | 0.0036 | 0.0036 | 0.0048 |
| ($C_{2v}$) | $Cu^+$ | 0.0008 | 0.0008 | -0.0048 |
| $CuAl_4^-$ | $Al_4^{2-}$ | 0.0178 | 0.0332 | 0.0154 |
| ($C_{4v}$) | $Cu^+$ | 0.0131 | -0.0023 | -0.0154 |

**TABLE 11: Profiles of the forward activation energy ($\Delta E_f^{\#}$), reverse activation energy ($\Delta E_r^{\#}$) and reaction energy ($\Delta E^0$) of a thermoneutral reaction ($F_a^- + CH_3\text{-}F_b \rightarrow F_a^-\text{-}CH_3 + F_b^-$; an endothermic reaction (HNO → HON) and an exothermic reaction ($H_2OO$ → HOOH).**

| Reaction | $\Delta E_f^{\#}$ | $\Delta E_r^{\#}$ | $\Delta E^0$ | $\xi_1$ | $\xi_2$ | *W1* | *W2* | *W3* | *W4* |
|---|---|---|---|---|---|---|---|---|---|
| **Thermo-neutral** B3LYP/6-311++G** | 9.54 | 9.54 | 0.0 | -1.33 | 1.33 | 5.42 | 4.12 | -4.12 | - 5.42 |
| **Endothermic** B3LYP/6-311+G** | 75.39 | 34.84 | 40.55 | -0.80 | 0.60 | 43.97 | 31.42 | -13,20 | - 21.64 |
| **Exothermic** B3LYP/6-311+G** | 7.39 | 52.85 | -45.46 | -0.65 | 0.87 | 3.93 | 3.46 | - 19.99 | - 32.86 |

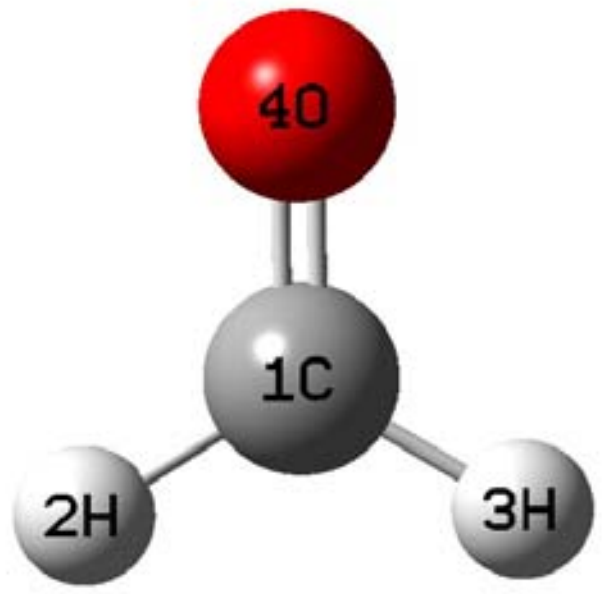

**a**

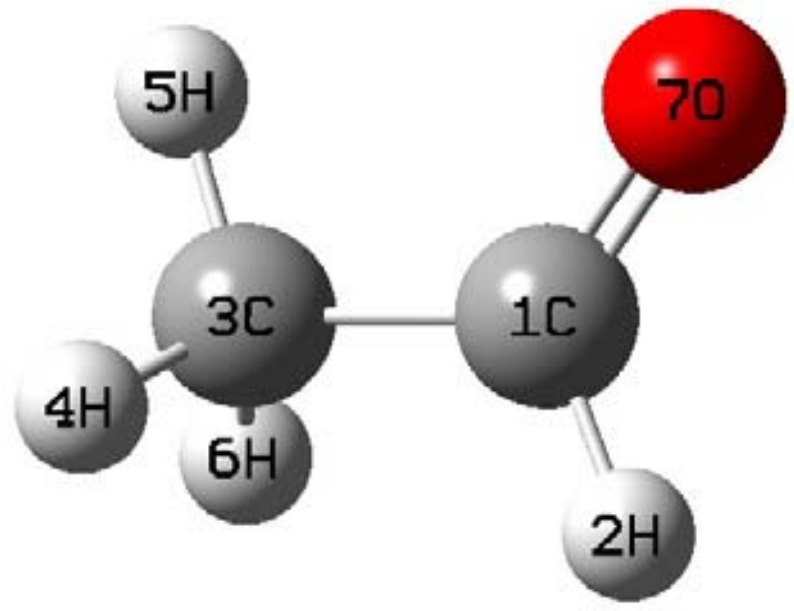

**b**

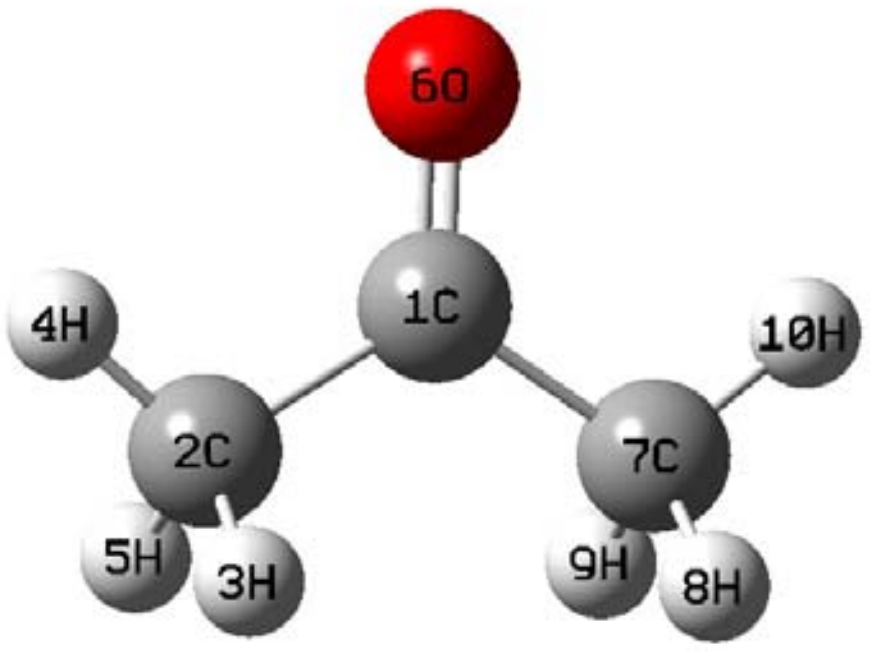

**c**

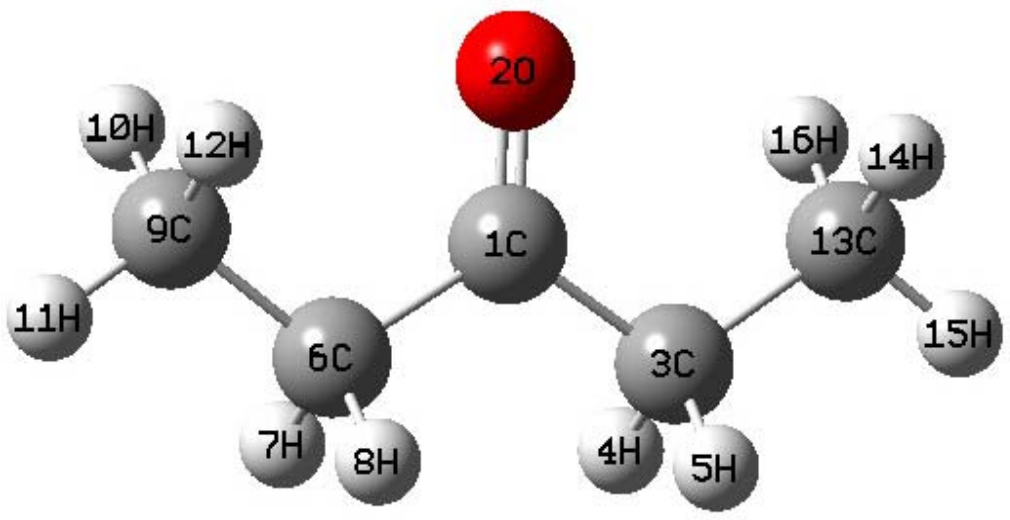

**d**

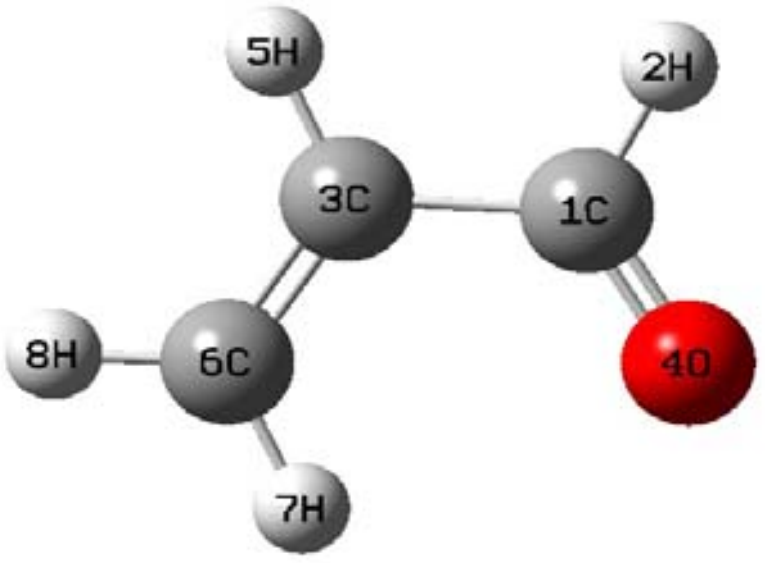

**e**

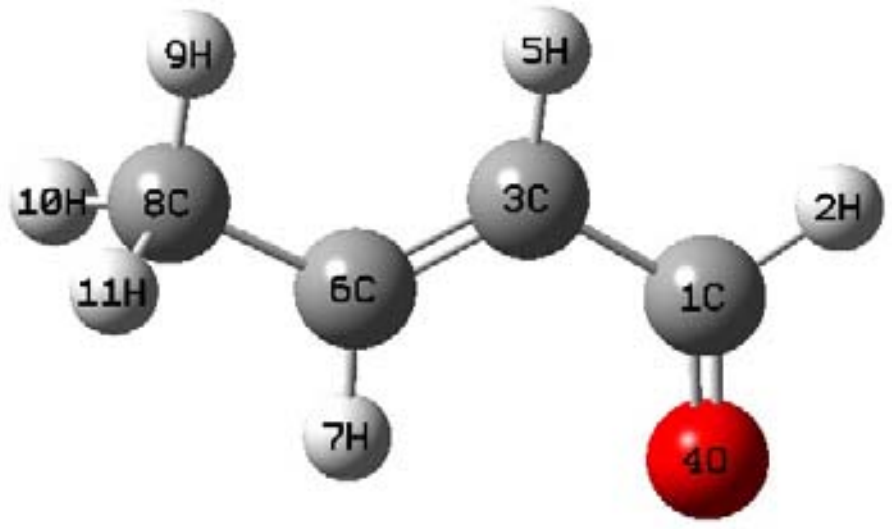

**f**

**Figure 1.** Optimized structures with atom numbering for the selected carbonyl compounds.

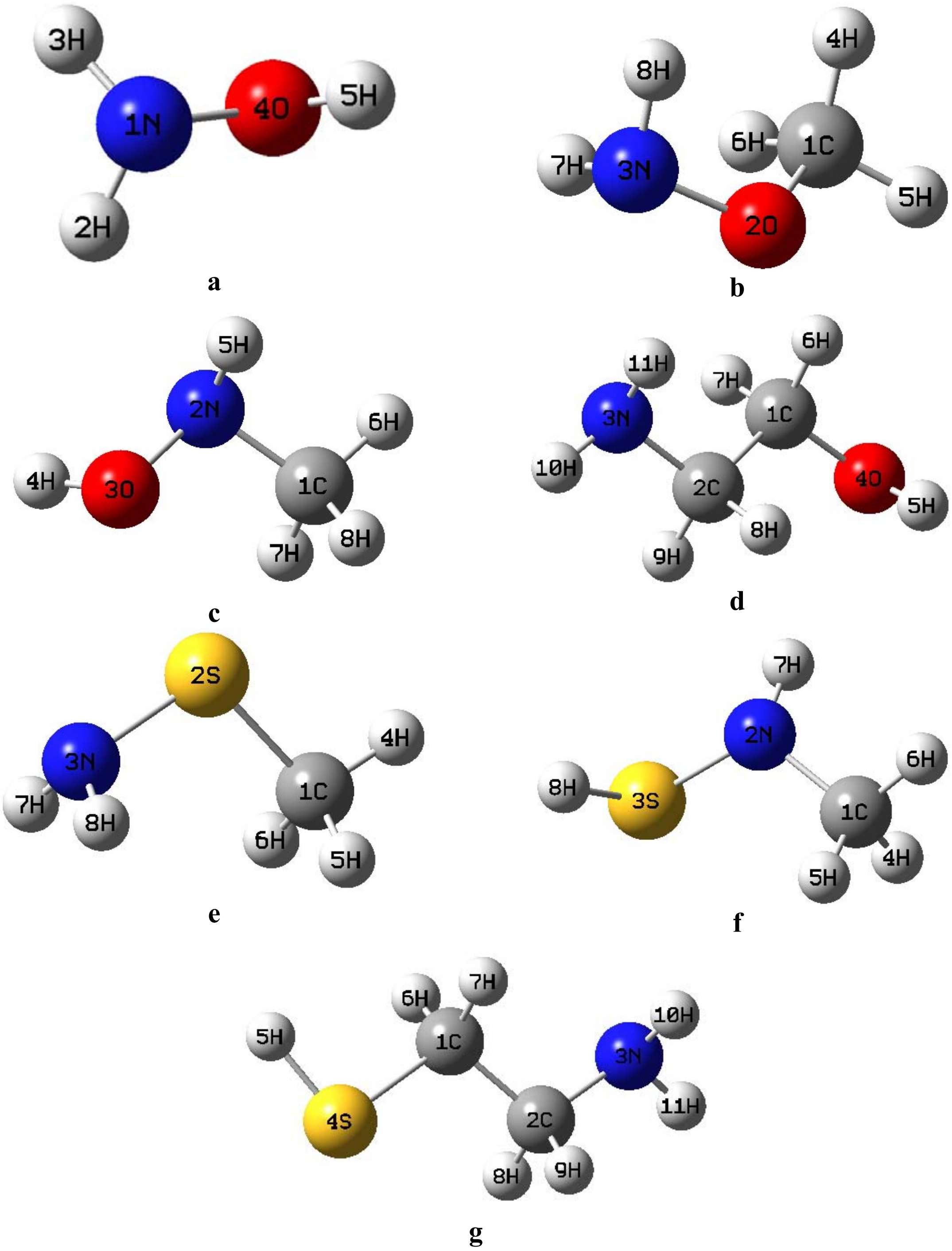

**Figure 2.** Optimized structures with atom numbering for the selected amine systems.

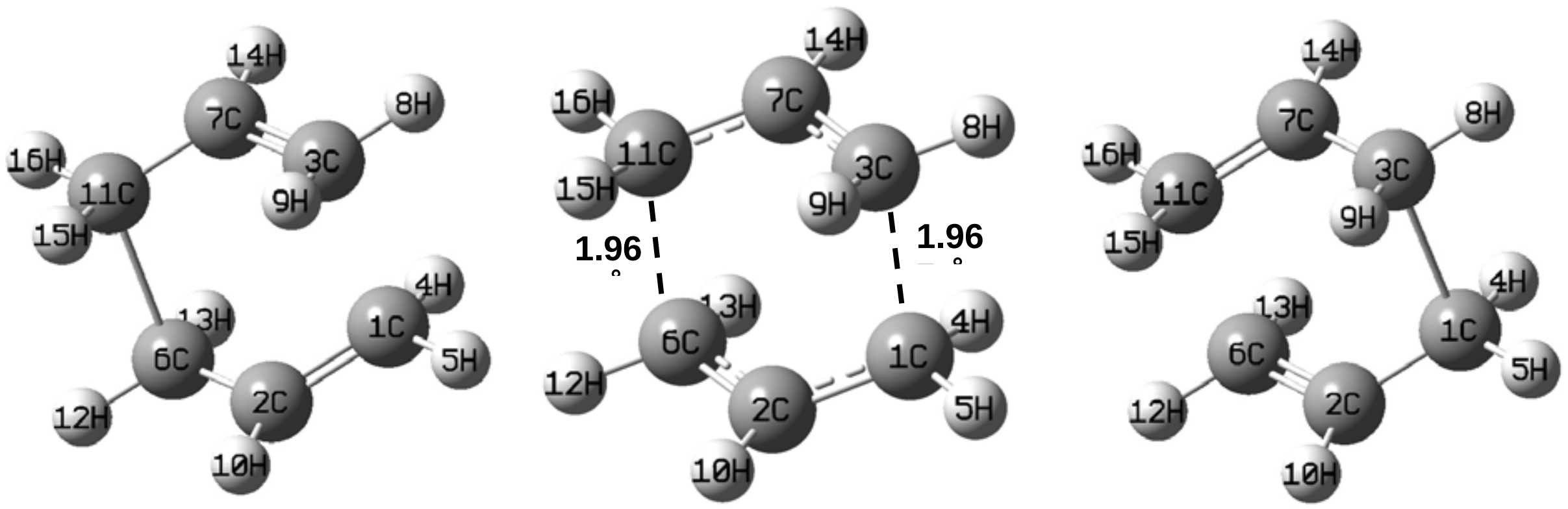

Reactant Transition State Product

**Figure 3:**Optimized geometrical structures calculated using B3LYP/6-31G* level of theory.

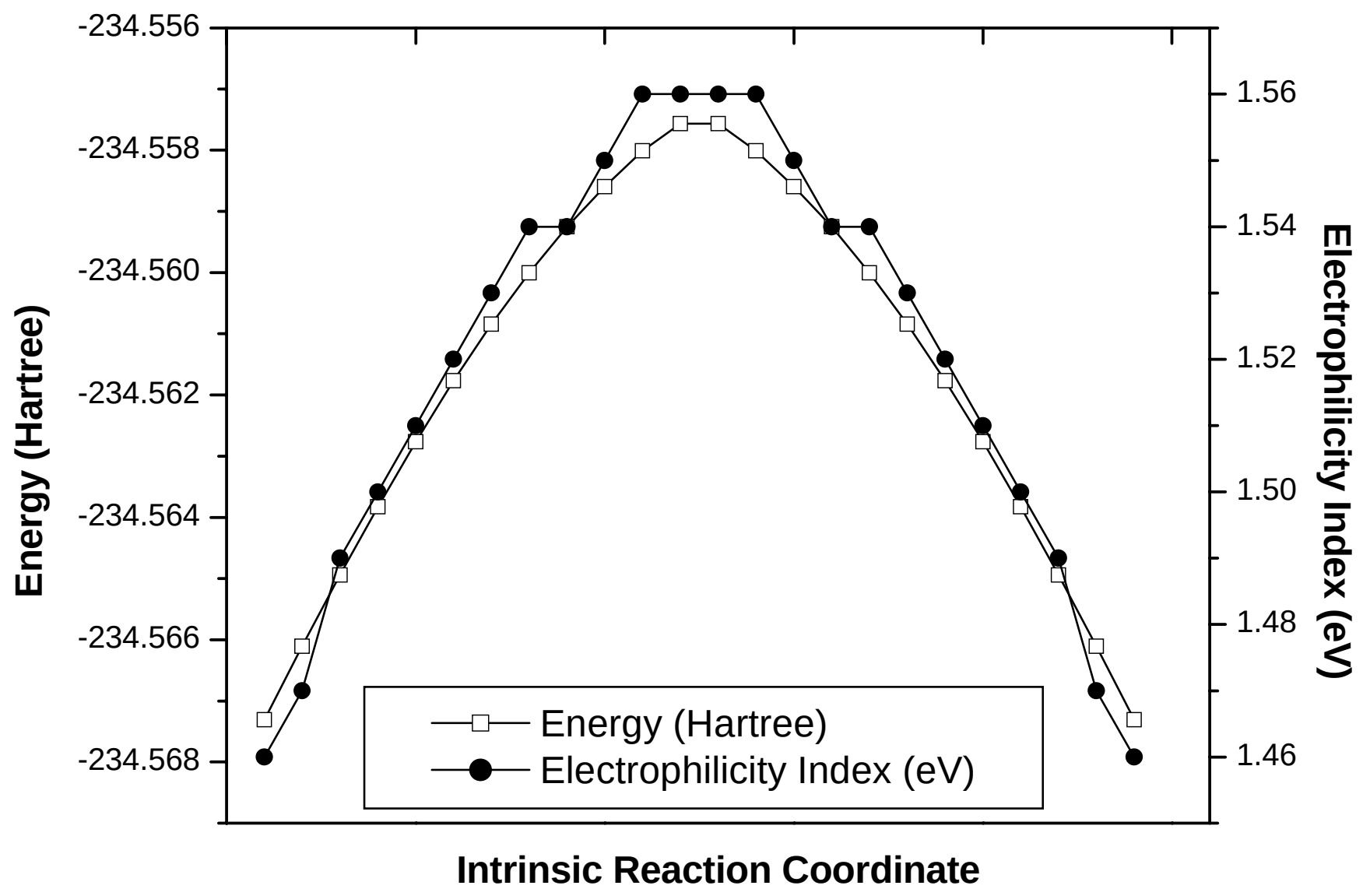

a

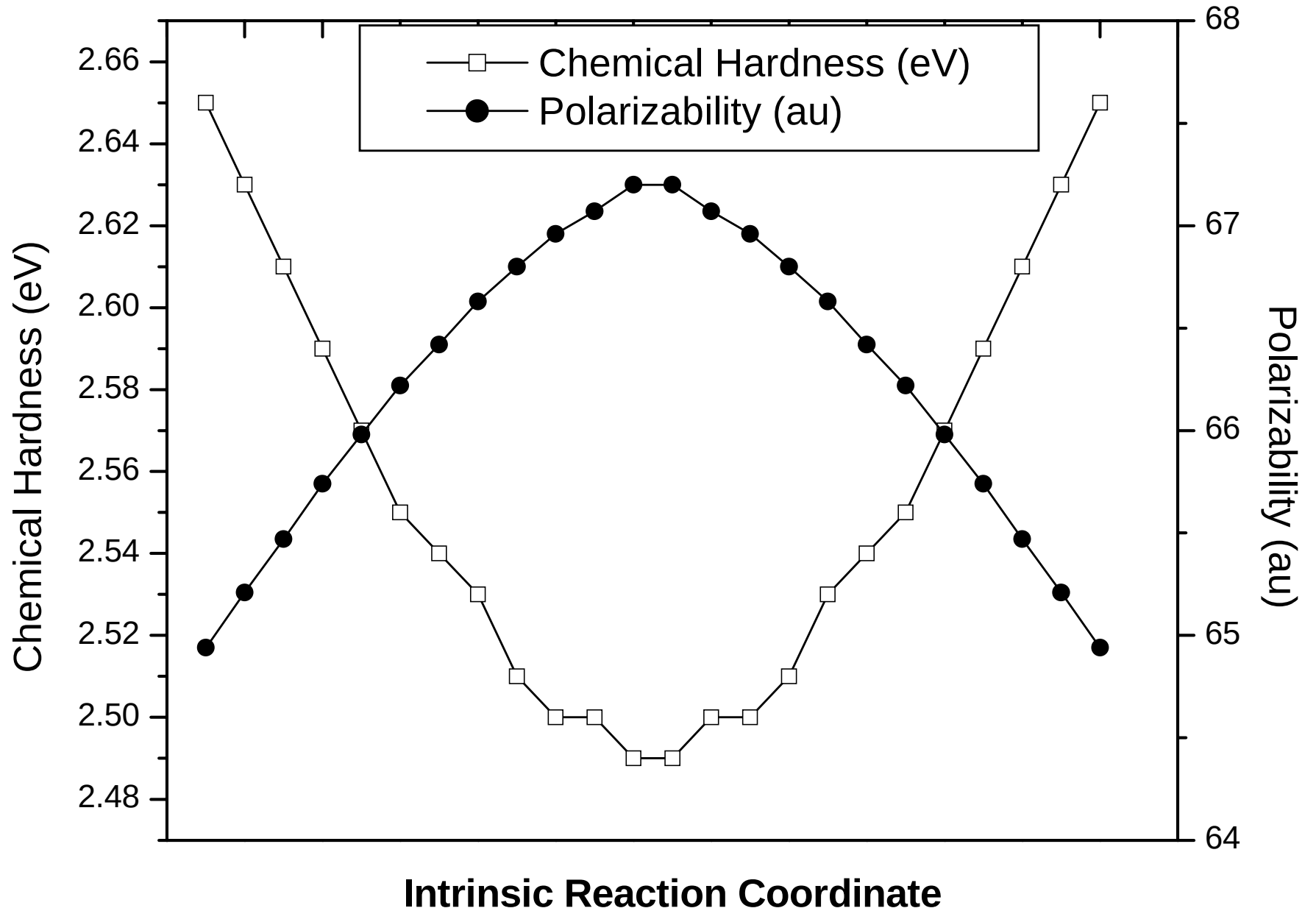

b

**Figure 4 (a-b):**Variation of global reactivity descriptors along intrinsic reaction coordinate.

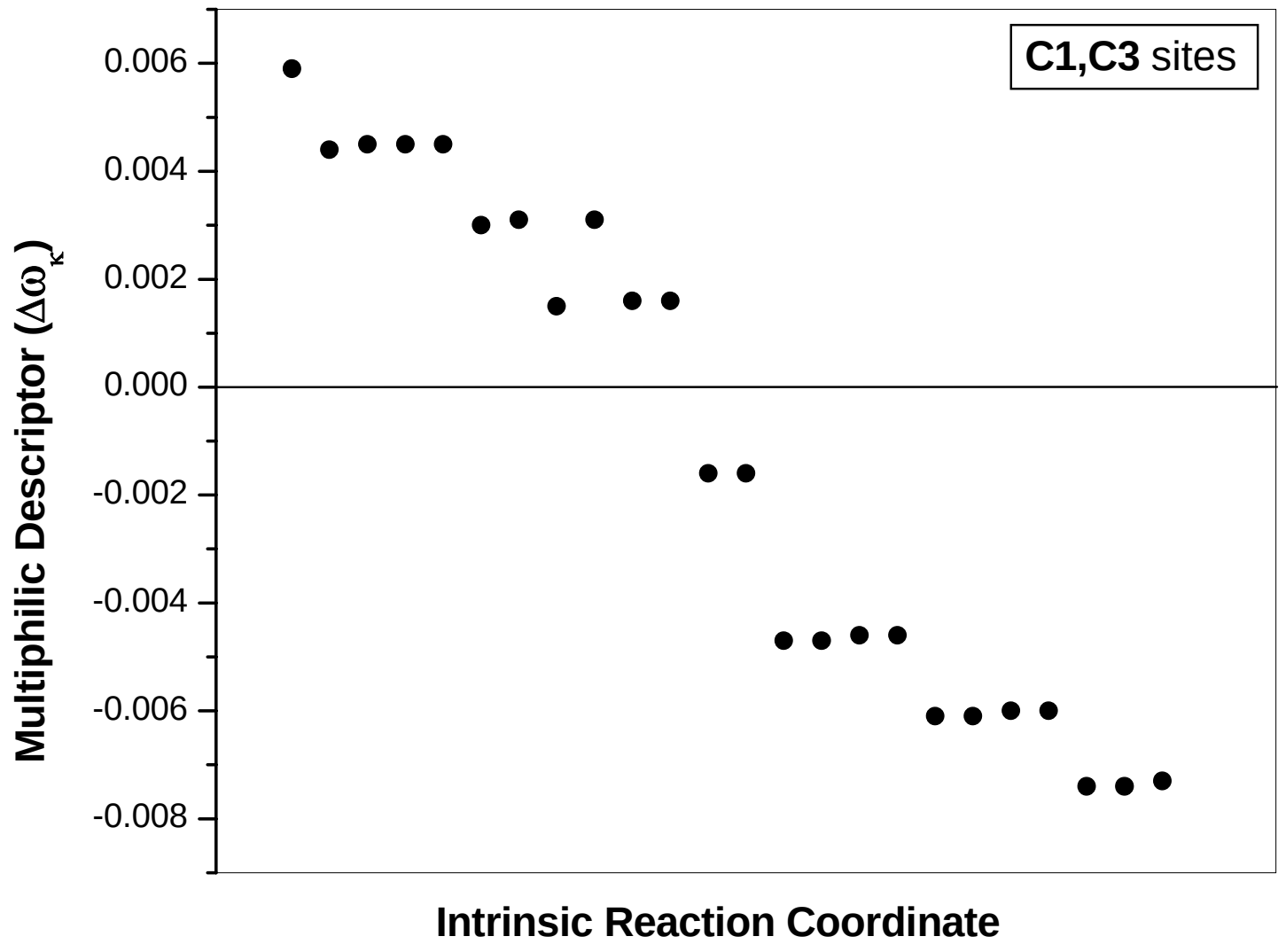

**a**

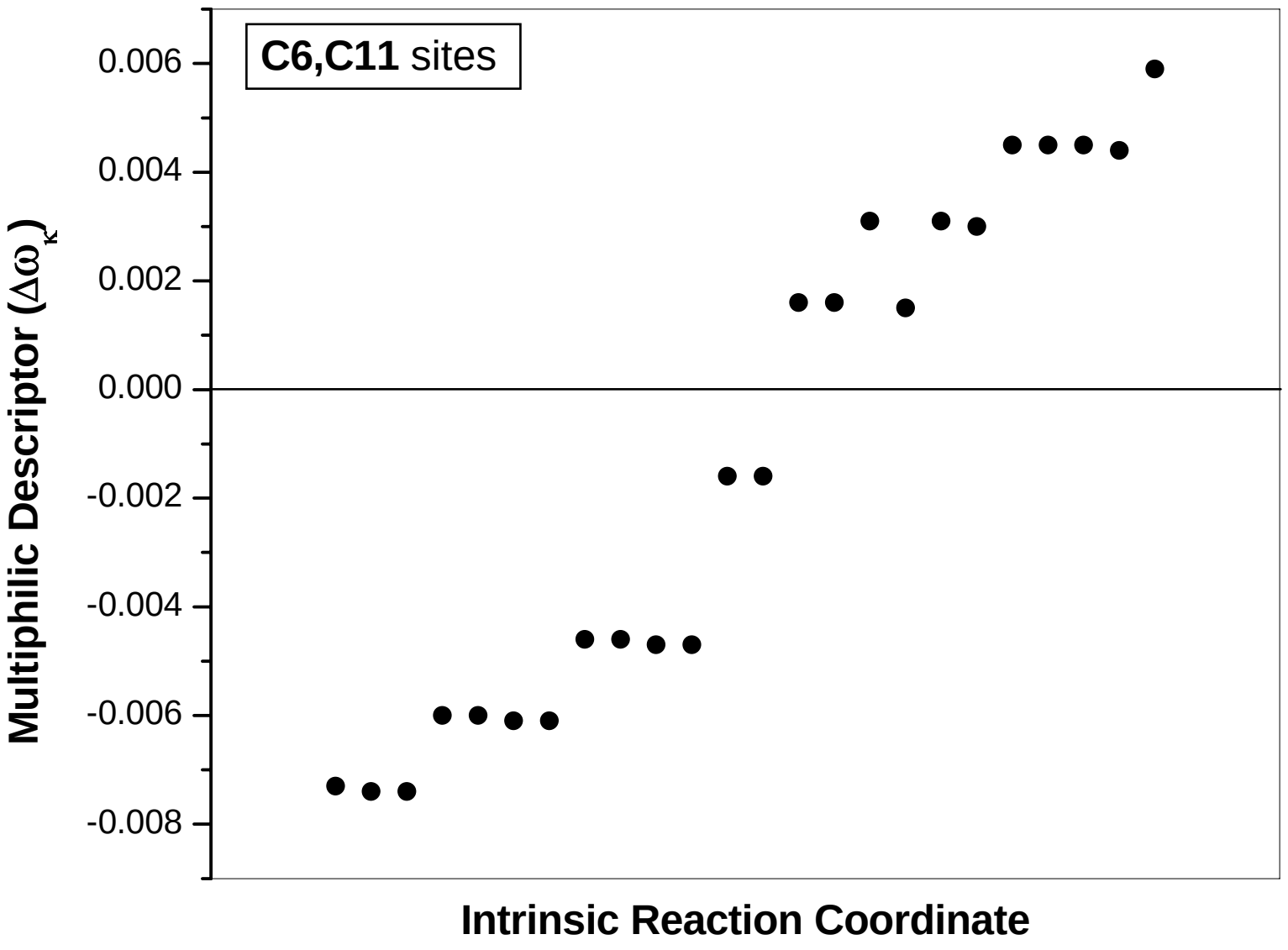

**b**

**Figure 5 (a-b):** Variation of multiphilic descriptor along intrinsic reaction coordinate for the selected atomic sites.

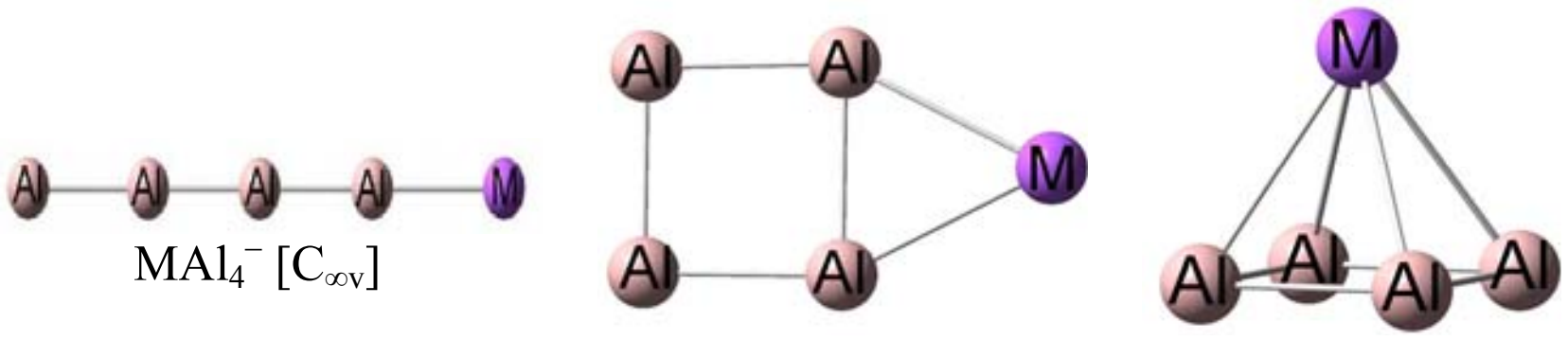

$MAl_4^-$ [$C_{\infty v}$]

$MAl_4^-$ [$C_{2v}$]

$MAl_4^-$ [$C_{4v}$]

M=Li, Na, K, Cu

**Figure 6.** Optimized structures of various isomers of $MAl_4^-$ (M ≡ Li, Na, K, Cu).

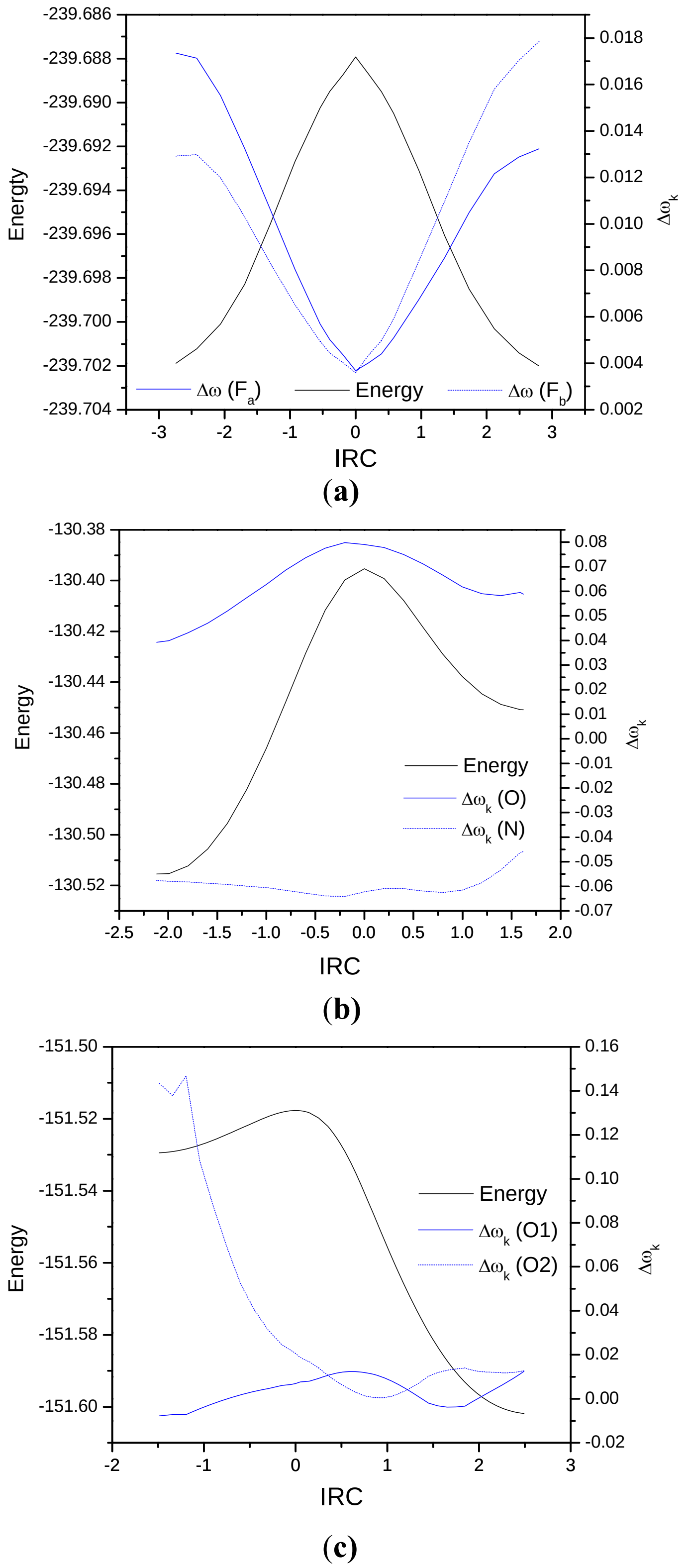

**Figure 7 (a-c):** Profiles of net nucleophilicity ($\Delta\omega_k$) of along the path of the gas phase (a) thermoneutral $S_N2$ substitution: $F_a^- + CH_3\text{-}F_b \rightarrow F_a\text{-}CH_3 + F_b^-$, (b) endothermic reaction: HNO → HON and (c) exothermic reaction: $H_2OO$ → HOOH. Also shown is the profile of energy.

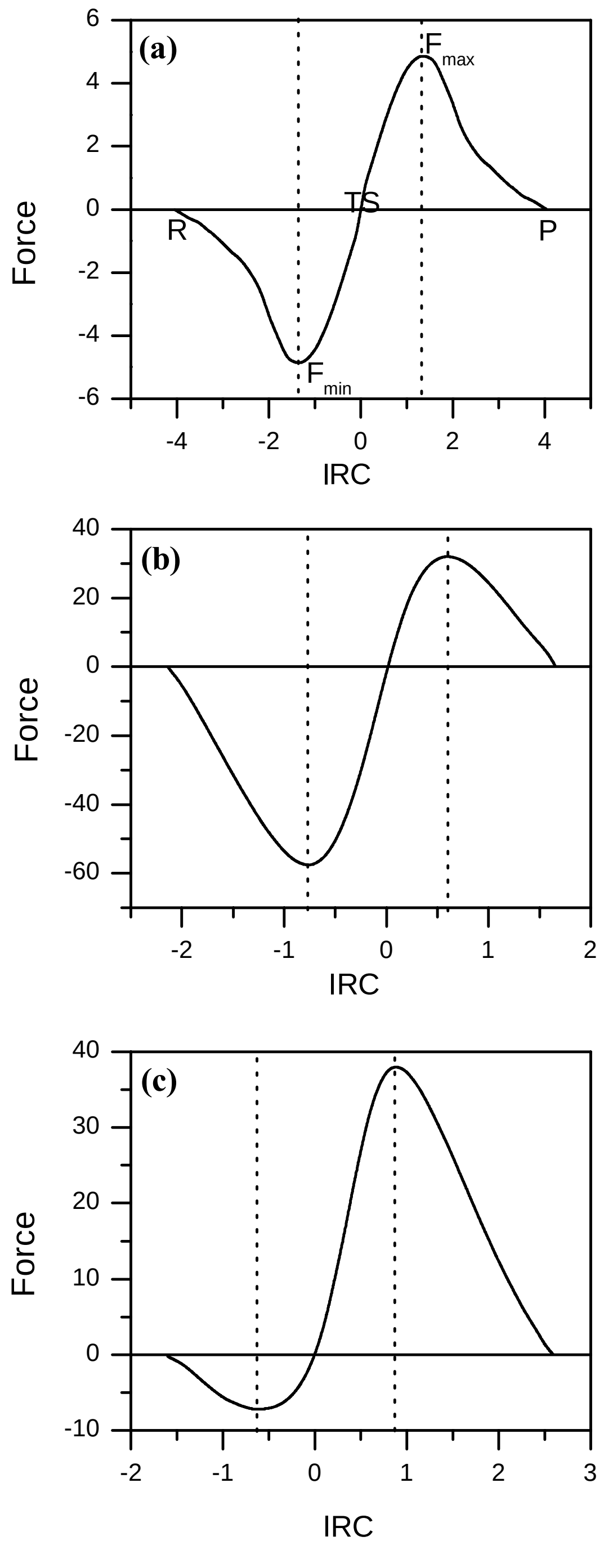

**Figure 8** Reaction force profiles along the reaction coordinate for **(a)** thermoneutral reaction: $F_a^- + CH_3\text{-}F_b \rightarrow F_a^-\text{-}CH_3 + F_b^-$; **(b)** endothermic reaction: HNO → HON; **(c)** the exothermic reaction: $H_2OO \rightarrow HOOH$. The vertical dashed lines define the reaction regions as follows: reactant (left), transition state (middle) and product (right).